\documentclass{ws-ijmpd}
\usepackage[super,compress]{cite}
\usepackage{amsmath,amsfonts,amssymb,mathrsfs,times,bm}
\usepackage{extarrows}
\usepackage{graphicx}
\usepackage{float}
\usepackage{graphicx,epstopdf}% Include figure files
\usepackage{dcolumn}% Align table columns on decimal point
\usepackage{exscale}
\usepackage{relsize}
\usepackage[colorlinks=true,breaklinks=true,linkcolor=blue,citecolor=blue,urlcolor=blue]{hyperref}
\newcommand{\nn}{\nonumber}

\begin{document}

\markboth{WEIJUN LI et al.}
{Exact Metric of a Constantly Moving Kerr Black Hole and Light Deflection}

\catchline{}{}{}{}{}
%
%%%%%%%%%%%%%%%%%%%%%%%%%%%%%%%%%%%%%%%%%%%%%%%%%%%%%%%%%%%%%%%%%%%%

\title{Kerr-Schild Form of the Exact Metric for a Constantly Moving Kerr Black Hole and Null Gravitational Deflection}

\author{WEIJUN LI}

\address{School of Mathematics and Physics, \\ University of South China, Hengyang, 421001, China}

\author{ZHONGWEN FENG}

\address{Physics and Space Science College, \\ China West Normal University, Nanchong, 637009, China}

\author{XIA ZHOU}

\address{Physics and Space Science College, \\ China West Normal University, Nanchong, 637009, China}

\author{XUELING MU}

\address{School of Information Science and Engineering, \\ Chengdu University, Chengdu, 610106, China}

\author{GUANSHENG HE\footnote{Corresponding author.}}

\address{School of Mathematics and Physics, \\ University of South China, Hengyang, 421001, China\\
hgs@usc.edu.cn}

\maketitle

%\begin{history}
%\received{Day Month Year}
%\revised{Day Month Year}
%\end{history}

\begin{abstract}
The exact metric of a moving Kerr black hole with an arbitrary constant velocity is derived in Kerr-Schild coordinates. We then calculate the null equatorial gravitational deflection caused by a radially moving Kerr source up to the second post-Minkowskian order, acting as an application of the weak field limit of the metric. The bending angle of light is found to be consistent with the result given in the previous works.
\end{abstract}


\ccode{PACS numbers: 95.30.Sf, 98.62.Sb}

%\tableofcontents

\section{Introduction}
The time dependence of a background field due to the translational motion of a gravitational system relative to the observer has an effect on the propagation of electromagnetic signals, which is the so-called velocity effect~\cite{PB1993,WS2004,HD2005}. Historically, this kinematical effect was usually ignored before the late stage of the last century, with the consideration of the low-velocity characteristic of most of the gravitational systems in the universe and the limited observational accuracy. In 1983, Birkinshaw and Gull\,~\cite{BG1983} found that the motion of a gravitational lens induces a characteristic two-sided pattern for the brightness of an isotropic radiation field, which in turn was proposed to measure the transverse velocity of a cluster of galaxies. Since this pioneering work, velocity effects have attracted more and more attentions, and the investigation of the kinematically correctional effects on the propagation of light or massive particles has been performed extensively (See, for instance,
Refs.~\citen{GM1986,KS1999,K2001,KM2002,Sereno2002,FK2003,Fr2003,MB2003,K2003,KP2003,Sereno2005,KF2007,KM2007,BAI2008,Kopei2009,ZKS2013,HBL2014,HL2014,SH2015,BZ2015,HL2016b,Zschocke2018,Zschocke2019}).
One of the main reasons responsible for its prosperity lies in the rapid progresses made in the techniques of position, velocity, and angular measurements, such as the achievement in the technique of the Very Long Baseline Interferometry (VLBI)~\cite{K2001,Burke1969,Rogers1970,TMS1986,Hirabayashi1998,Ma1998} and the ones in the subsequent techniques applied in current high-accuracy astrometry projects (e.g., the Square Kilometer Array (SKA) Project~\cite{CR2004,DHSL2009} and the Gaia mission~\cite{Perryman2001,Mignard2007,Prusti2016}).

As is well known, the study of the motion effects in gravitational physics depends fundamentally on the achievement of the metric of a moving gravitational body, for which two common methods are feasible.
One is the retarded-potential-solution approach~\cite{KS1999,KM2002}, where the Li\'{e}nard-Wiechert gravitational potential acts as the solution of the Einstein field equations. The other is the Lorentz boosting technique which is based on the existing solutions of the field equations~\cite{PB1993,Weinberg1972}. To our knowledge, the first work to obtain the weak-field time-dependent metric of a slowly moving point mass by means of the latter method was accomplished in Ref.~\citen{PB1993}. In 2004, Wucknitz and Sperhake~\cite{WS2004} made use of the Lorentz boosting technique to get the first post-Minkowskian (PM) metric of a constantly moving body, and investigated the velocity effects on the leading-order gravitational deflection of light and massive particles. In fact, probing strong-field gravitational physics also plays a important role in the examinations of general relativity (GR) and alternative theories of gravitation~\cite{Will2014,Psaltis2008}, which calls for the exact time-dependent metric of the central body. Recently, on the basis of the Lorentz transformation method~\cite{PB1993}, the exact metrics for moving Schwarzschild\,~\cite{HL2014a} and Reissner-Nordstr\"{o}m~\cite{HL2014b} black holes with a constant velocity have been obtained in harmonic coordinates. However, it is found that the calculations of the strict metric of a moving Kerr black hole performed in harmonic coordinates become very complex. This issue has not been dealt with so far, and is worth the effort, since the case of a black hole with an intrinsic angular momentum~\cite{Kerr1963,ZLY2015} and thus the motion effects related to it is more realistic than that for a black hole without rotation.

In this paper, we apply the Lorentz boosting technique to derive the exact metric of a moving Kerr black hole with an arbitrary constant velocity, serving as our main result. The derivations are performed in Kerr-Schild coordinates~\cite{Kerr1963,BL1967}, which have two following advantages: (1) the Kerr-Schild form avoids the coordinate singularities and it may make the calculations more convenient; (2) it is suitable to compute the physical parameters of the metric, as demonstrated in Ref.~\citen{Kerr2007}. An additional motivation to use Kerr-Schild coordinates lies in the fact that the exact analytical solution of the Einstein's field equation for a moving rotating black hole in standard coordinates is too lengthy to be applied conveniently, although the metrics of static or stationary black holes in Boyer-Linquist coordinates are elegant and have been adopted by lots of work (See, for instance, Refs.~\citen{Font1999,Campa2001,CLG2012,Bhatt2020}).

It should be pointed out that the solution derived in this paper may have good prospect in potential astrophysical applications (e.g., gravitational physics~\cite{Weinberg1972,Bozza2008,Br2018,Wei2019,Barack2019,He2020} and spacetime properties~\cite{Virbhadra1990,XW2017,GPS2020}). As we know, rotating black holes have been verified recently by the most direct evidences~\cite{Abbott2016,Akiyama2019}, while it is widely believed that all existing astrophysical black holes can be described by the Kerr solution~\cite{ZLY2015}. What's more, the relative motions of the source of light, the observer, and the central body are common in the universe. There is no doubt that moving rotating black holes with a translational velocity should not be rare. For example, the final black hole resulted from the merger of two unequal-mass black holes receives a ``kick" and has a recoil velocity and a spin, due to the asymmetric loss of linear momentum in the gravitational radiation~\cite{GSBHH2007}. If the translational velocity of a kicked rotating black hole is constant or the acceleration of a moving primordial Kerr source is too small to be taken into account, the gravitational field of the black hole can be described by our exact metric.

We know the gravitational lensing phenomenon is one of the most classical tests of various gravitational theories including GR (See, e.g., Refs.~\citen{Wam1998,Perlick2004,AR2002,CJ2009,WLFY2012,CXZZ2015,LYJ2016,ZX2016,ZX2017a,ZX2017b,HL2017,JWBO2017,Jusufi2018,OJS2018,CX2018,JOSVG2018,JO2018,PJ2019,WSX2019,LX2019,ZX2020,JKAO2020}, and references therein). As an application of the weak field limit of our exact metric in Kerr-Schild form, we calculate the equatorial gravitational deflection of light due to a radially moving Kerr black hole up to the 2PM order, in the weak-field and small-angle approximation. For simplicity, the source of light and the observer are assumed to be static relative to the rest frame of the background.

The organization of this paper is as follows. Section~\ref{sect2} presents the derivation of the exact metric of a moving Kerr black hole with a constant velocity. Section~\ref{sect3} is devoted to calculating the null gravitational deflection caused by a radially moving Kerr black hole in the second post-Minkowskian approximation. A summary is given in Section~\ref{sect4}. Throughout the paper, the geometrized units where $G = c = 1$ and the metric signature $(-,~+,~+,~+)$ are used. Conventionally, Greek indices run over $0,~1,~2$, and $3$, while Latin indices run over $1,~2$, and $3$.

\section{The Kerr-Schild form for the metric of a moving Kerr source} \label{sect2}
We adopt the Lorentz boosting technique~\cite{PB1993,WS2004} to achieve the metric of a constantly moving Kerr black hole. Let $\{\bm{e}_1,~\bm{e}_2,~\bm{e}_3\}$ be the orthonormal basis of a three-dimensional Cartesian coordinate system. We assume the rest Kerr-Schild frame of the background and the comoving Kerr-Schild frame for the barycenter of the gravitational source to be $(t,~x,~y,~z)$ and $(T,~X,~Y,~Z)$, respectively.
The Kerr-Schild form of the Kerr metric in the comoving frame $(T,~X,~Y,~Z)$ can be then written as~\cite{Kerr1963,BL1967,DKS1969,KS2009}
\begin{eqnarray}
\nn&&ds^2=-\,dT^2+dX^2+dY^2+dZ^2+\frac{2MR^3}{R^4+a^2Z^2}  \\
&&\hspace*{28pt}\times\left[dT+\frac{ZdZ}{R}+\frac{R\,(XdX+YdY)-a\,(XdY-YdX)}{R^2+a^2}\right]^2 ,  \label{metric-1}
\end{eqnarray}
where $\frac{X^2+Y^2}{R^2+a^2}+\frac{Z^2}{R^2}=1$, and $M$ and $a\equiv J/M$ denote the rest mass and angular momentum per mass of the gravitational source, with $\bm{J} (=J\bm{e}_3)$ being its intrinsic angular momentum vector along $z$-axis.

In order to obtain the line element in the background's rest frame $(t,~x,~y,~z)$, in which the Kerr black hole is moving with an arbitrary constant velocity $\bm{v}=v_1\bm{e}_1+v_2\bm{e}_2+v_3\bm{e}_3$,
we apply a Lorentz boost to the metric presented in Eq.~(\ref{metric-1}). The Lorentz transformation between $(t,~x,~y,~z)$ and the comoving frame $(T,~X,~Y,~Z)$ is given by
\begin{equation}
X^\mu = \Lambda^\mu_\nu x^\nu~, \label{LorentzTran0}
\end{equation}
and
\begin{equation}
\Lambda^\mu_\nu=\left(\begin{array}{cccc}
    \gamma & -v_1\gamma & -v_2\gamma & -v_3\gamma \vspace*{2pt}  \\
    -v_1\gamma & 1+\frac{v_1^2(\gamma-1)}{v^2} & \frac{v_1v_2(\gamma-1)}{v^2} & \frac{v_1v_3(\gamma-1)}{v^2} \vspace*{2pt} \\
    -v_2\gamma & \frac{v_1v_2(\gamma-1)}{v^2} & 1+\frac{v_2^2(\gamma-1)}{v^2} & \frac{v_2v_3(\gamma-1)}{v^2} \vspace*{2pt} \\
    -v_3\gamma & \frac{v_1v_3(\gamma-1)}{v^2} & \frac{v_2v_3(\gamma-1)}{v^2} & 1+\frac{v_3^2(\gamma-1)}{v^2} \vspace*{2pt}
                        \end{array} \right)~,~
\end{equation}
where $\gamma= (1-v^2)^{-\scriptstyle \frac{1}{2}}$ is Lorentz factor and $v^2=v_1^2+v_2^2+v_3^2$. The Kerr-Schild form of the exact metric for the arbitrarily constantly moving Kerr black hole can be obtained as follows:
%\begin{widetext}
\begin{eqnarray}
   &&g_{00}=-1+\frac{2\gamma^2MR^3}{R^4+a^2Z^2}\!\left[1-\frac{v_1(RX+aY)}{R^2+a^2}-\frac{v_2(RY-aX)}{R^2+a^2}-\frac{v_3Z}{R}\right]^2~,       \label{g00mK}     \\
\nn&&g_{0i}=\frac{2\gamma MR^3}{R^4+a^2Z^2}\left[1-\frac{v_1(RX+aY)}{R^2+a^2}-\frac{v_2(RY-aX)}{R^2+a^2}-\frac{v_3Z}{R}\right]       \\
\nn&&\hspace*{27pt}\times\left\{\frac{(RX\!+aY)\delta_{1i}}{R^2+a^2}+\frac{(RY\!-aX)\delta_{2i}}{R^2+a^2}+\frac{Z\delta_{3i}}{R} \right. \\
   &&\left.\hspace*{25.5pt} +\,\frac{v_i(\gamma-1)}{v^2}\!\left[\frac{v_1(RX+aY)}{R^2+a^2}+\frac{v_2(RY-aX)}{R^2+a^2}+\frac{v_3Z}{R}\right]\!-v_i\gamma\right\}~,   \label{g0imK}    \\
\nn&&g_{ij}=\delta_{ij}+\frac{2MR^3}{R^4+a^2Z^2}\!\left\{\frac{(RX\!+aY)\delta_{1i}}{R^2+a^2}+\frac{(RY\!-aX)\delta_{2i}}{R^2+a^2}+\frac{Z\delta_{3i}}{R} \right. \\
\nn&&\left.\hspace*{23pt} +\,\frac{v_i(\gamma-1)}{v^2}\!\left[\frac{v_1(RX+aY)}{R^2+a^2}+\frac{v_2(RY-aX)}{R^2+a^2}+\frac{v_3Z}{R}\right]\!-v_i\gamma\right\}  \\
\nn&&\hspace*{24pt}\times\left\{\frac{(RX+aY)\delta_{1j}}{R^2+a^2}+\frac{(RY-aX)\delta_{2j}}{R^2+a^2}+\frac{Z\delta_{3j}}{R} \right. \\
   &&\left.\hspace*{23pt}+\,\frac{v_j(\gamma-1)}{v^2}\!\left[\frac{v_1(RX+aY)}{R^2+a^2}+\frac{v_2(RY-aX)}{R^2+a^2}+\frac{v_3Z}{R}\right]\!-v_j\gamma\right\}~.~~~~~~  \label{gijmK}
\end{eqnarray}
%\end{widetext}

In the limit $a\rightarrow0$, Eqs.~(\ref{g00mK}) - (\ref{gijmK}) are reduced to the exact metric of a moving Schwarzschild black hole with an arbitrary constant velocity in Kerr-Schild coordinates:
\begin{eqnarray}
&&g_{00}=-1+\frac{2\gamma^2M}{R}\left(1-\frac{\bm{v}\cdot\!\bm{X}}{R}\right)^2~,       \label{g00mS}     \\
&&g_{0i}=\frac{2\gamma M}{R}\!\left(\!1\!-\!\frac{\bm{v}\cdot\!\bm{X}}{R}\!\right)\!\!\left[\frac{X_i}{R}\!-\!v_i\gamma\!+\!\frac{v_i(\gamma\!-\!1)(\bm{v}\cdot\!\bm{X})}{v^2R}\right]~,~~~~\label{g0imS}    \\
\nn&&g_{ij}=\delta_{ij}+\frac{2M}{R}\left[\frac{X_i}{R}-v_i\gamma+\frac{v_i(\gamma-1)(\bm{v}\cdot\!\bm{X})}{v^2R}\right]    \\
&&\hspace*{23pt}\times\left[\frac{X_j}{R}-v_j\gamma+\frac{v_j(\gamma-1)(\bm{v}\cdot\!\bm{X})}{v^2R}\right]~,  \label{gijmS}
\end{eqnarray}
with $R=\sqrt{X^2+Y^2+Z^2}$ and $\bm{X}=(X,~Y,~Z)$. It can be seen that when $\bm{v}=0$, Eqs.~(\ref{g00mS}) - (\ref{gijmS}) are further simplified to the Schwarzschild metric presented in Kerr-Schild coordinates in the background's rest frame, which reads~\cite{KS2009,Edd1924}:
\begin{eqnarray}
&&g_{00}=-1+\frac{2M}{r}~,                    \label{g00S}     \\
&&g_{0i}=\frac{2Mx_i}{r^2}~,                  \label{g0iS}     \\
&&g_{ij}=\delta_{ij}+\frac{2Mx_ix_j}{r^3}~,   \label{gijS}
\end{eqnarray}
where $r=\sqrt{x^2+y^2+z^2}$.

\section{Light deflection by a radially moving Kerr source up to the 2PM order} \label{sect3}
As an application of the weak field limit of the metric given in Eqs.~(\ref{g00mK}) - (\ref{gijmK}), we consider the 2PM gravitational deflection of light induced by a moving Kerr black hole with a radial velocity~\cite{WS2004}. Note that for the convenience of discussion to be followed, we still denote the comoving Kerr-Schild frame of the radially moving source as $(T,~X,~Y,~Z)$.

\subsection{The metric in the weak field limit}
The metric of a radially moving Kerr source with $\bm{v}=v_1\bm{e}_1=v\bm{e}_1$ up to the 2PM order in Kerr-Schild coordinates can be obtained from
Eqs.~(\ref{g00mK}) - (\ref{gijmK}) as follows:
\begin{eqnarray}
&&g_{00}=-1+\frac{2\gamma^2M}{R}\left(1-\frac{vX}{R}\right)\left[1-v\left(\frac{X}{R}+\frac{2aY}{R^2}\right)\right]+O(M^3)~,             \label{g00mS-WF}   \\
\nn&&g_{0i}=\frac{2\gamma M}{R}\!\left\{\left(1-\frac{vX}{R}-\frac{v\hspace*{1pt}aY}{R^2}\right)\!\left[\frac{X_i}{R}+v\gamma\left(\frac{v\gamma}{\gamma+1}\frac{X}{R}-1\right)\delta_{i1}\right] \right. \\
&&\left.\hspace*{24pt}+\left(1-\frac{vX}{R}\right)\frac{a(\gamma Y\delta_{i1}-X\delta_{i2})}{R^2}\right\}+O(M^3)~,   \label{g0imS-WF}   \\
\nn&&g_{ij}=\delta_{ij}\!+\!\frac{2M}{R}\!\left\{\left[\frac{X_i}{R}\!+\!\frac{a(Y\delta_{i1}\!-\!X\delta_{i2})}{R^2}
\!+\!v\gamma\left(\frac{v\gamma}{\gamma+1}\!\left(\frac{X}{R}\!+\!\frac{aY}{R^2}\right)\!-\!1\right)\delta_{i1} \right] \right. \\
\nn&&\left.\hspace*{24pt}\times\left[\frac{X_j}{R}+\left(\frac{(\gamma-1)X}{R}-v\gamma\right)\delta_{j1}\right]
+\left[\frac{X_i}{R}+\left(\frac{(\gamma-1)X}{R}-v\gamma\right)\delta_{i1}\right] \right. \\
&&\left.\hspace*{24pt}\times \frac{a(\gamma Y\delta_{j1}-X\delta_{j2})}{R^2} \right\}+O(M^3) ~, ~~~~~~         \label{gijmS-WF}
\end{eqnarray}
for which the Lorentz transformation~(\ref{LorentzTran0}) is reduced to
\begin{eqnarray}
&& T=\gamma(t-vx)~,   \label{LT-T}   \\
&& X=\gamma(x-vt)~,   \label{LT-X}   \\
&& Y=y~,              \label{LT-Y}   \\
&& Z=z~,              \label{LT-Z}
\end{eqnarray}
with $R=\sqrt{\gamma^2(x-vt)^2+y^2+z^2}$. The elements of the inverse of this weak-field metric tensor up to the 1PM order are
{\small
\begin{eqnarray}
&&\hspace*{-0.2cm}g^{00}=-1\!-\!\frac{2\gamma^2M(R\!-\!vX)^2}{R^3}\!+\!O(M^2)~,~~~~~~ g^{11}=1\!-\!\frac{2\gamma^2M(X\!-\!vR)^2}{R^3}\!+\!O(M^2)~,~~~~~~~~ \label{g00mS-WF-I}   \\
&&\hspace*{-0.2cm}g^{22}=1-\frac{2MY^2}{R^3}+O(M^2)~,         \hspace*{1.85cm}        g^{33}=1-\frac{2MZ^2}{R^3}+O(M^2)~,       \label{g0imS-WF-I}   \\
&&\hspace*{-0.2cm}g^{01}=\frac{2\gamma^2M\!\left[(1\!+\!v^2)RX\!-\!v(R^2\!+\!X^2)\right]}{R^3}\!+\!O(M^2)~, ~~~g^{02}=\frac{2\gamma M(R\!-\!vX)Y}{R^3}\!+\!O(M^2)~,~~~~~~~ \label{gijmS-WF-I}  \\
&&\hspace*{-0.2cm}g^{03}=\frac{2\gamma M(R-vX)Z}{R^3}+O(M^2)~,\hspace*{1.05cm}        g^{12}=\frac{2\gamma M(vR-X)Y}{R^3}+O(M^2)~,~~~~  \\
&&\hspace*{-0.2cm}g^{13}=\frac{2\gamma M(vR-X)Z}{R^3}+O(M^2)~,\hspace*{1.05cm}        g^{23}=-\frac{2MYZ}{R^3}+O(M^2)~.
\end{eqnarray}}

\subsection{Equatorial equations of motion}
Via calculating the Christoffel symbols, the null geodesic equations constrained to the equatorial plane ($z=\frac{\partial}{\partial z}=0$) of the lens up to the 2PM order can be obtained in the background's rest frame:
%\begin{widetext}
{\small
\begin{eqnarray}
\nn&&\hspace*{-0.2cm}0=\ddot{t}+\!\gamma^3\Bigg{\{}\!\frac{M\!\left\{2v^2(X^2\!-\!y^2)R\!-\!v\!\left[R^2\!+\!v^2(X^2\!-\!2y^2)\right]\!X\!\right\}}{R^5}\!+\!\frac{2v^2Ma\!\left[3XR\!+\!v(y^2\!-\!3X^2)\right]\!y}{R^6}  \\
\nn&&\hspace*{0.3cm}+\,\frac{2M^2\left[(1+3v^2)X^4-v\,(3+v^2)X^3R+(2+3v^2)X^2y^2-3vXy^2R+y^4\right]}{R^7}\Bigg{\}}\,\dot{t}^2  \\
\nn&&\hspace*{0.3cm}+\,\gamma^3\Bigg{\{}\frac{M\left[2(X^2-y^2)R-vX(3X^2-v^2R^2)\right]}{R^5}+\frac{2Ma\left[3XR+v(y^2-3X^2)\right]y}{R^6}  \\
\nn&&\hspace*{0.3cm}+\frac{2M^2\left[(1+3v^2)X^4-v(3\!+\!v^2)X^3R+(1\!+\!4v^2)X^2y^2-v(2\!+\!v^2)Xy^2R+v^2y^4\right]}{R^7}\Bigg{\}}\,\dot{x}^2  \\
\nn&&\hspace*{0.3cm}+\,2\,\gamma^3\Bigg{\{}\!\frac{M\!\left[(1\!+\!v^2)X^3\!+\!(1\!-\!2v^2)Xy^2\!+\!2v(y^2\!-\!X^2)R\right]}{R^5}\!-\!\frac{2vMa\!\left[3XR\!+\!v(y^2\!-\!3X^2)\right]\!y}{R^6}  \\
\nn&&\hspace*{0.3cm}+\frac{2M^2\left\{X\!\left[R^2+v^2(3X^2+2y^2)\right]-vR\left[(3+v^2)X^2+y^2\right] \right\}}{R^6}\Bigg{\}}\,\dot{t}\,\dot{x}  \\
   &&\hspace*{0.3cm}+\frac{2\gamma^2M(R^2\!-\!4vXR\!+\!3v^2X^2)\,y}{R^5}\,\dot{t}\,\dot{y}+\!\frac{2\gamma^2M\!\left[4XR\!-\!v(4X^2\!+\!y^2)\right]\!y}{R^5}\dot{x}\,\dot{y}+O(M^3)~,   \label{ME-t}
\end{eqnarray}}
{\small
\begin{eqnarray}
\nn&&\hspace*{-0.2cm}0=\ddot{x}+\gamma^3\Bigg{\{}\!\frac{M\!\left[(1\!-\!3v^2)X^3+Xy^2+2v^3(X^2\!-\!y^2)R\right]}{R^5}\!+\!\frac{2v^2Ma\left(y^2-3X^2+3vXR\right)y}{R^6}  \\
\nn&&\hspace*{0.3cm}-\frac{2M^2\left\{XR\left[R^2+v^2(3X^2+2y^2)\right]-vR^2\left[(3+v^2)X^2+y^2\right]\right\}}{R^7}\Bigg{\}}\,\dot{t}^2  \\
\nn&&\hspace*{0.3cm}+\,\gamma^3\Bigg{\{}\frac{M\!\left[(2\!-\!v^2)Xy^2-(1\!+\!v^2)X^3+2v(X^2\!-\!y^2)R\right]}{R^5}\!+\!\frac{2Ma\left(y^2\!-\!3X^2+3vXR\right)y}{R^6}  \\
\nn&&\hspace*{0.3cm}-\frac{2M^2\left[XR(X^2+3v^2R^2)-vR^2(3X^2+v^2R^2)\right]}{R^7}\Bigg{\}}\,\dot{x}^2+\,2\gamma^3  \\
\nn&&\hspace*{0.3cm}\times\Bigg{\{}\!\frac{vM\!\left\{\!X\!\left[(1\!+\!v^2)X^2\!-\!(2\!-\!v^2)y^2\right]\!-\!2v(X^2\!-\!y^2)R\right\}}{R^5}\!-\!\frac{2vMa\left(y^2\!-\!3X^2\!+\!3vXR\right)y}{R^6}   \\
\nn&&\hspace*{0.3cm}-\frac{2M^2\!\left[(1\!+\!3v^2)X^4\!-\!v(3\!+\!v^2)X^3R\!+\!(1+4v^2)X^2y^2\!-\!v(2+v^2)Xy^2R\!+\!v^2y^4\right]}{R^7}\!\Bigg{\}}\,\dot{t}\,\dot{x}   \\
\nn&&\hspace*{0.3cm}+\frac{2v\gamma^2M(4X^2+y^2-4vXR)\,y}{R^5}\,\dot{t}\,\dot{y}-\frac{2\gamma^2M\!\left[(3+v^2)X^2+v^2y^2-4vXR\right]y}{R^5}\,\dot{x}\,\dot{y} \\
   &&\hspace*{0.3cm}+\,O(M^3)~,    \label{ME-x}  \\
\nn&&\hspace*{-0.2cm}0=\ddot{y}+\!\Bigg{\{}\!\frac{M\!\left[X^2\!+\!(3\gamma^2\!-\!2)y^2\right]\!y}{R^5}
\!-\!\frac{v\gamma^2Ma\!\left[2R^3\!-\!2vX(X^2\!-\!3y^2)\right]}{R^6}\!-\!\frac{2\gamma^2M^2(R\!-\!vX)^2y}{R^6}\!\Bigg{\}}\,\dot{t}^2   \\
\nn&&\hspace*{0.27cm}-\Bigg{[}\frac{M(R^2-3\gamma^2y^2)\,y}{R^5}-\frac{2\gamma^2Ma\,(X^3-3Xy^2-vR^3)}{R^6}+\frac{2\gamma^2M^2(X-vR)^2\,y}{R^6}\Bigg{]}\,\dot{x}^2-2\gamma^2  \\
\nn&&\hspace*{0.27cm}\times\Bigg{\{}\!\frac{3vMy^3}{R^5}\!-\!\frac{Ma\!\left[(1\!+\!v^2)R^3\!+\!2vX(3y^2\!-\!X^2)\right]}{R^6}
\!+\!\frac{2M^2\!\!\left[(1\!+\!v^2)XR\!-\!v(2X^2\!+\!y^2)\right]\!y}{R^6}\!\Bigg{\}}\,\dot{t}\,\dot{x}   \\
   &&\hspace*{0.27cm}+\frac{6v\gamma MXy^2}{R^5}\,\dot{t}\,\dot{y}\!-\!\frac{6\gamma MXy^2}{R^5}\,\dot{x}\,\dot{y}+O(M^3)~,  \label{ME-y}
\end{eqnarray}}
%\end{widetext}
where $R=\sqrt{X^2+Y^2}$, and a dot denotes the derivative with respect to the affine parameter $\xi$ along the geodesic~\cite{WS2004,Weinberg1972}. Since $\dot{y}$ is related to the gravitational deflection angle by Eq.~(\ref{angle-1}) below and thus regarded to be of the first order of magnitude, the terms with the factor $\dot{y}^2$ on the right-hand side of Eqs.~(\ref{ME-t}) - (\ref{ME-y}) belong to third- or higher-order terms and have been omitted.

\subsection{2PM gravitational deflection by a radially moving Kerr source} \label{3.1.3}
\begin{figure*}[t]
\centering
\includegraphics[width=12.8cm]{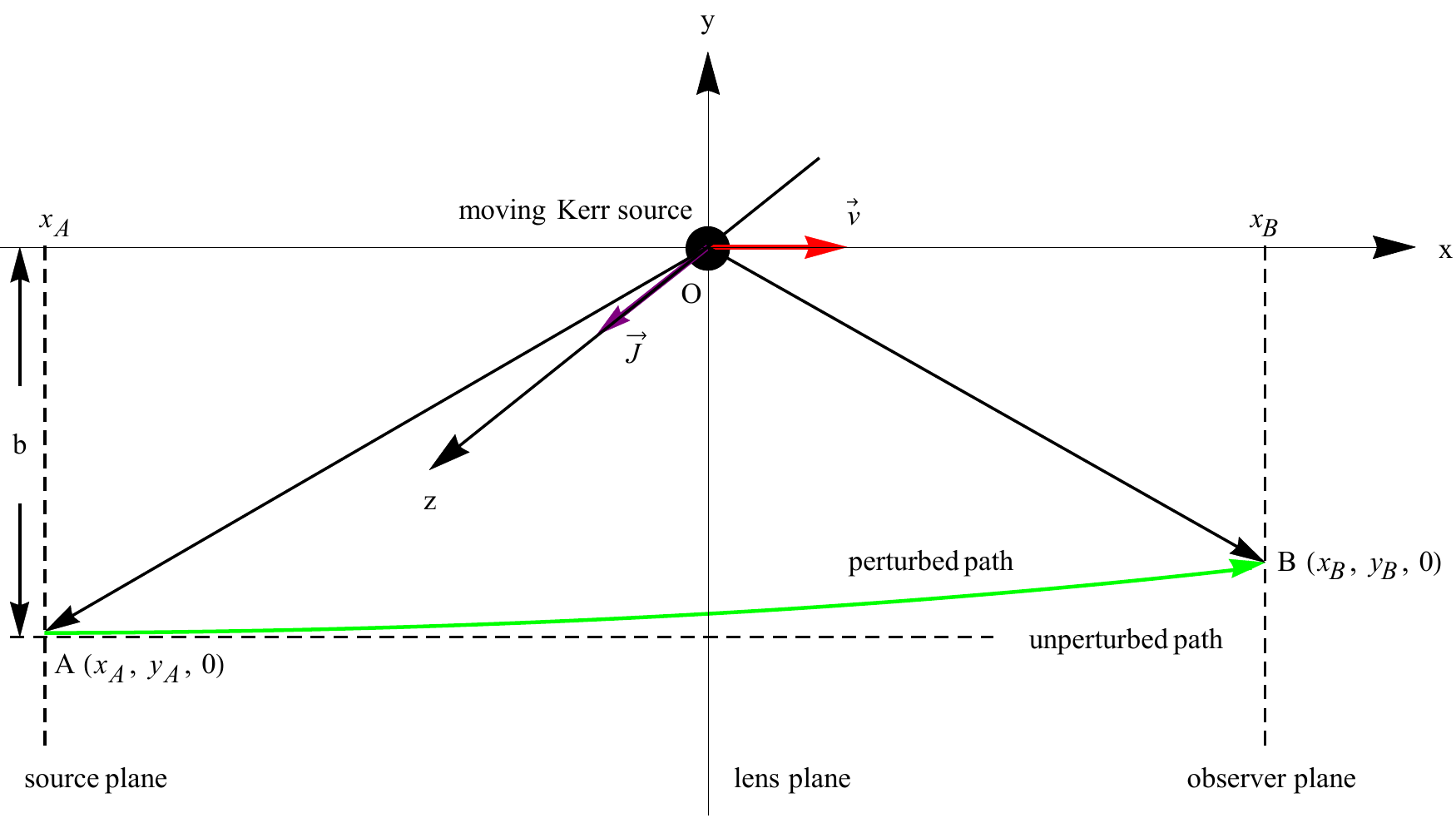}
\caption{The geometrical diagram for the equatorial propagation of light in the gravitational field of a radially moving Kerr black hole. The gravitational deflection is greatly exaggerated to distinguish the perturbed path (green) from the unperturbed one (dashed horizontal). }   \label{Figure1}
\end{figure*}

The geometrical diagram for light propagating in the equatorial plane of a radially moving Kerr black hole is shown in Fig.~\ref{Figure1}. The spatial coordinates of the source (denoted as $A$) and observer (denoted as $B$) of light in the background's rest Kerr-Schild frame are assumed as $(x_A,~y_A,~0)$ and $(x_B,~y_B,~0)$, respectively, where $y_A\approx-b$, $x_A\ll-b$, and $x_B\gg b$, with $b$ being the impact parameter. The spatial coordinates of the emission and reception events in the comoving frame are given respectively to be $(X_A,~Y_A,~0)$ and $(X_B,~Y_B,~0)$. The green line denotes the perturbed propagation path of light coming from $x\rightarrow-\infty$ with an initial velocity $\bm{w}|_{x\rightarrow -\infty}=\bm{e}_1$.

In the following calculations of light bending, we adopt an iterative technique, and assume the path parameter $\xi$ to have the dimension of length, as done in Refs.~\citen{WS2004,HL2017}. The gravitational deflection angle of a test particle propagating from $\xi\rightarrow -\infty$ to $\xi\rightarrow +\infty$ is defined by
\begin{eqnarray}
\nn&&\alpha\equiv\arctan\left. \frac{dy}{dx}\right|_{\xi\rightarrow +\infty}-\arctan \left.\frac{dy}{dx}\right|_{\xi\rightarrow -\infty} \\
&&\hspace*{9pt}=\arctan\left. \frac{\dot{y}}{\dot{x}}\right|_{x\rightarrow +\infty}-\arctan \left.\frac{\dot{y}}{\dot{x}}\right|_{x\rightarrow -\infty}~, ~~~     \label{angle-1}
\end{eqnarray}
which can also be written approximately as under the scenario where both the emitter and the observer are far away from the lens with a finite distance~\cite{HL2017}
\begin{eqnarray}
\alpha\approx\arctan\left. \frac{\dot{y}}{\dot{x}}\right|_B-\arctan \left.\frac{\dot{y}}{\dot{x}}\right|_A~.      \label{angle-2}
\end{eqnarray}

Up to the zero order, Eqs.~(\ref{ME-t}) - (\ref{ME-y}) yield
\begin{eqnarray}
&& \dot{t}=1+O(M)~,       \label{0PM-dott}  \\
&& \dot{x}=1+O(M)~,       \label{0PM-dotx}  \\
&& \dot{y}=0+O(M)~,       \label{0PM-doty}
\end{eqnarray}
where we have used the boundary conditions $\dot{t}|_{\xi\rightarrow -\infty}=\dot{t}|_{x\rightarrow -\infty}=1$, $\dot{x}|_{\xi\rightarrow -\infty}=\dot{x}|_{x\rightarrow -\infty}=1$, and $\dot{y}|_{\xi\rightarrow -\infty}=\dot{y}|_{x\rightarrow -\infty}=0$. Eq.~(\ref{0PM-dotx}) gives a 0PM parameter transformation
\begin{equation}
dx=\left[1+O(M)\right]d\xi~.     \label{0PM-PT}
\end{equation}
Moreover, based on the boundary condition $y|_{\xi\rightarrow -\infty}=y|_{x\rightarrow -\infty}=-b$, Eq.~(\ref{0PM-doty}) leads to the zero-order form of $y$ as
\begin{equation}
y=-b+O(M)~.                      \label{0PM-y}
\end{equation}
In addition, the combination of Eqs.~(\ref{0PM-dott}) - (\ref{0PM-dotx}) and (\ref{LT-X}) gives the zero-order coordinate transformation
\begin{equation}
dX=\left[(1-v)\gamma+O(M)\right]dx~.      \label{LT-R-0PM}
\end{equation}

The substitution of Eqs.~(\ref{0PM-dott}) - (\ref{LT-R-0PM}) into Eqs.~(\ref{ME-t}) - (\ref{ME-y}) yields the explicit forms of $\dot{t},~\dot{x},$ and $\dot{y}$ up to the 1PM order, which read:
\begin{eqnarray}
&&\dot{t}=1+\frac{M\!\left[2\left(X^2+b^2\right)+2X\sqrt{X^2+b^2}+vb^2\right]}{(1+v)\left(X^2+b^2\right)^{\frac{3}{2}}}+\,O(M^2)~,               \label{1PM-dott}  \\
&&\dot{x}=1+\frac{M\!\left[b^2+2\,v\sqrt{X^2+b^2}\left(\sqrt{X^2+b^2}+X\right)\right]}{(1+v)\left(X^2+b^2\right)^{\frac{3}{2}}}+\,O(M^2)~,       \label{1PM-dotx}  \\
&&\dot{y}=\frac{(1-v)\gamma MX\left(2X^2+3b^2\right)}{b\left(X^2+b^2\right)^{\frac{3}{2}}}+O(M^2)~.       \label{1PM-doty}
\end{eqnarray}
Based on Eqs.~(\ref{0PM-PT}) and (\ref{LT-R-0PM}), the integration of Eq.~(\ref{1PM-doty}) over $\xi$ gives the 1PM form of $y$
\begin{equation}
y=-b\left[1-\frac{M(2X^2+b^2)}{b^2\sqrt{X^2+b^2}}\right]+O(M^2)~.         \label{1PM-y}
\end{equation}
Eq.~(\ref{1PM-dotx}) leads to the 1PM parameter transformation
\begin{equation}
d\xi=\left\{1-\frac{M\!\left[b^2+2\,v\sqrt{X^2+b^2}\left(\sqrt{X^2+b^2}+X\right)\right]}{(1+v)\left(X^2+b^2\right)^{\frac{3}{2}}}+O(M^2)\right\}dx~.     \label{1PM-PT}
\end{equation}
With the help of Eqs.~(\ref{LT-X}), (\ref{1PM-dott}) and (\ref{1PM-dotx}), we can also obtain the coordinate transformation up to the 1PM order
\begin{equation}
dx=\frac{1}{(1-v)\gamma}\left\{1+ \frac{vM\!\left[b^2+2X\left(\sqrt{X^2+b^2}+X\right)\right]}{(1+v)\left(X^2+b^2\right)^{\frac{3}{2}}}+O(M^2)\right\}dX~.      \label{LT-R-1PM}
\end{equation}
Substituting Eqs.~(\ref{1PM-dott}) - (\ref{LT-R-1PM}) into the integration of Eq.~(\ref{ME-y}) over $\xi$, we get the explicit form of $\dot{y}$ up to the 2PM order as follows:
{\small\begin{eqnarray}
\nn&&\hspace*{-0.2cm}\dot{y}=(1-v)^2\gamma^2\int\Bigg{\{}\frac{3Mb^3}{\left(X^2+b^2\right)^{\frac{5}{2}}}+\frac{3M^2b\left(2X^2+b^2\right)\left(4b^2-X^2\right)}{\left(X^2+b^2\right)^4}  \\
\nn&&\hspace*{4.35cm}-\frac{2Ma\!\left[X^3-3b^2X+(X^2+b^2)^{\frac{3}{2}}\right]}{\left(X^2+b^2\right)^3}\Bigg{\}}\,d\xi+O(M^3)  \\
\nn&&\hspace*{0.02cm}=(1-v)\gamma\int\Bigg{\{}\frac{3Mb^3}{\left(X^2+b^2\right)^{\frac{5}{2}}}+\frac{M^2b\left(9b^6+30b^4X^2+15b^2X^4-6X^6\right)}{\left(X^2+b^2\right)^5}  \\
\nn&&\hspace*{4.cm}+\,Ma\!\left[\frac{6b^2X-2X^3}{\left(X^2+b^2\right)^3}-\frac{2}{\left(X^2+b^2\right)^{\frac{3}{2}}}\right]\!\Bigg{\}}\,dX+O(M^3)  \\
\nn&&\hspace*{0.02cm}=(1-v)\gamma\Bigg{\{}\frac{M(2X^2+3b^2)X}{b\left(X^2+b^2\right)^{\frac{3}{2}}}
\!+\!\frac{3M^2\!\left[5bX^5+16b^3X^3+7b^5X+5(X^2+b^2)^3\arctan{\frac{X}{b}}\right]}{4b^2\left(X^2+b^2\right)^3}  \\
   &&\hspace*{0.36cm}-\frac{Ma\!\left[2X^5+b^2X^2\left(4X-\sqrt{X^2+b^2}\right)+b^4\left(2X+\sqrt{X^2+b^2}\right)\right]}{b^2\left(X^2+b^2\right)^{\frac{5}{2}}} \Bigg{\}}+O(M^3)~.       \label{2PM-doty}
\end{eqnarray}}

Finally, substituting Eqs.~(\ref{1PM-dotx}) and (\ref{2PM-doty}) into Eq.~(\ref{angle-2}), we achieve the explicit form of the null gravitational deflection angle up to the 2PM order in the background's rest Kerr-Schild frame as
\begin{equation}
\alpha=(1-v)\gamma\left(\frac{4M}{b}+\frac{15\pi M^2}{4b^2}-\frac{4Ma}{b^2}\right)+O(M^3)~,                      \label{2PM-angle}
\end{equation}
which matches well with the result obtained in the previous works (See, e.g., Refs.~\citen{WS2004,Sereno2005,KM2007,HL2016b}). Here, the weak-field conditions $\frac{M}{\sqrt{X_A^2+b^2}}\ll1$ and $\frac{M}{\sqrt{X_B^2+b^2}}\ll1$ have been used in the derivations. The contribution of the spin term on the right-hand side of Eq.~(\ref{2PM-angle}) is negative and positive for the prograde and retrograde motions of the photon relative to the rotation of the black hole, respectively. Note that the radial velocity effect on the second-order gravitational deflection of light and its measurability have been discussed in the previous work~\cite{HL2016b}. In addition, the consistency between Eq.~(\ref{2PM-angle}) and Eq.~(20) obtained in harmonic coordinates in Ref.~\citen{HL2017} (along with Eq.~(32) in Ref.~\citen{HL2016b}) indicates further the fact that observable physical quantities are independent on the concrete coordinates one uses.

At to Eq.~(\ref{2PM-angle}), a simple numerical example in astrophysical scenarios is given as follows. Let us consider the 2PM gravitational deflection of light caused
by a kicked Kerr black hole resulted from the merger of two unequal-mass black holes mentioned above~\cite{GSBHH2007}. The constant radial velocity and the spin of the black hole are assumed as $v=5.8\times10^{-4}$ (corresponding to a kick velocity $\sim 175.0$ km/s) and $a=0.5M$, respectively. The impact parameter is set to be $b=1.0\times10^5M$.
Then, the first-order mass-induced, second-order mass-induced and spin-induced contributions to the total deflection angle $\alpha$ in Eq.~(\ref{2PM-angle}) can be evaluated to be about
$8.3$\,as, $242.9\,\mu$as, and $41.2\,\mu$as (absolute value), respectively.

\section{Summary} \label{sect4}
In this paper, we have worked out the exact metric of a moving Kerr black hole with an arbitrary constant velocity in Kerr-Schild coordinates via the Lorentz boosting technique. As an application of the weak field limit of this solution, the null gravitational deflection in the equatorial plane of a radially moving Kerr black hole up to the second post-Minkowskian order has been computed, in accordance with the result presented in the previous works. Considering the commonness of the relative motion of the source of light, the observer, and the deflector, this new solution of the Einstein field equations might be helpful to study the strong- and weak-field physics of a moving gravitational system.

\section*{Acknowledgments}
We would like to thank the referee for his/her valuable comments that have improved this manuscript. This work was supported by the National Natural Science Foundation of China (Grant Nos. 11947018 and 11947128).

\end{document}